\documentclass[twocolumn,aps,prd,amssymb,eqsecnum,amsfonts,epsf,preprintnumbers,nofootinbib,superscriptaddress]{revtex4}
\usepackage[colorlinks,
            linkcolor=blue,
            anchorcolor=blue,
            citecolor=green
            ]{hyperref}
\usepackage{amsmath,amsfonts,latexsym,amssymb,graphics,epsfig,subfigure,color,makeidx}

\begin{document}
\title{Braneworld in $f(T)$ Gravity Theory with Noncanonical Scalar Matter Field}
\author{Jian Wang\footnote{wangjian16@lzu.edu.cn}}
\author{Wen-Di Guo\footnote{guowd14@lzu.edu.cn}}
\author{Zi-Chao Lin\footnote{linzch16@lzu.edu.cn}}
\author{Yu-Xiao Liu\footnote{liuyx@lzu.edu.cn, corresponding author}}

\affiliation{Research Center of Gravitation $\&$ Institute of Theoretical Physics, Lanzhou University, Lanzhou 730000, China}
\affiliation{Key Laboratory for Magnetism and Magnetic of the Ministry of Education, Lanzhou University, Lanzhou 730000, China}

\begin{abstract}
In this paper, we investigate the braneworld scenario in $f(T)$ gravity with a $K$-field as the background field. We consider various different specific forms of $f(T)$ gravity and $K$-field, and find a general way to construct the braneworld model. Based on our solutions, the split of branes is investigated. Besides, the stability of the braneworld is studied by investigating the tensor perturbation of the vielbein.
\end{abstract}

\maketitle

\section{Introduction}
One of the most well-known and earliest extra dimension theories was first proposed by T. Kaluza \cite{Kaluza:1921tu} and O. Klein \cite{Klein:1926fj} to unify Einstein's general
relativity and Maxwell's electromagnetism in the 1920s. The extra dimension theories drew wide attention with the work of N.
Arkani-Hamed, S. Dimopoulos, and G. R. Dvali \cite{ArkaniHamed:1998rs} and the works of L. Randall and R. Sundrum \cite{Randall:1999ee,Randall:1999vf} in the end of the 20th century. Later, various braneworld scenarios were developed such as the Gregory-Rubakov-Sibiryakov (GRS) model \cite{Gregory:2000jc}, the Dvali-Gabadadze-Porrati (DGP) model \cite{Dvali:2000hr}, the thick brane model \cite{DeWolfe:1999cp,Gremm:1999pj,Csaki:2000fc,Gremm:2000dj,Dzhunushaliev:2009yx,Dzhunushaliev:2009va,HerreraAguilar:2009wc,Dzhunushaliev:2008zz,Gogberashvili:2003ys,Gogberashvili:2003xa}, the universal extra dimension model\cite{Appelquist:2000nn}, etc.
Among these theories, one important model is the thick brane theory originated from the domain wall model proposed by V. A. Rubakov and M. E. Shaposhnikov \cite{Rubakov:1983bb} in 1983. In the thick brane model, the brane could be generated by scalar fields \cite{Kehagias2001a,Andrianov:2005hm,Gremm:2000dj,Afonso:2006gi,Bazeia:2006ef,Bazeia:2002sd,Neupane:2010ey,Dutra:2014xla}, as well as vector fields and spinor fields \cite{Geng:2015kvs,Dzhunushaliev:2009yx,Dzhunushaliev:2011mm}. In addition, there are also braneworld models without matter fields \cite{zhong:2015pta,HerreraAguilar:2009wc,liu:2011am}. The standard model fields in the bulk can be localized near the brane \cite{Gregory:2000jc,Dvali:1996xe,Gremm:2000dj,German:2012rv,Liu:2008wd,Liu:2009dt}.

The braneworld was also studied in different modified gravity theories, for example, the scalar-tensor gravity theory \cite{Bogdanos2006,Bogdanos2007,Yang2012,Farakos2007,Herrera-Aguilar2012,Gogberashvili2013,German2014,Liu2012,Guo2012,Karam2018}, the metric $f(R)$ gravity theory \cite{Parry:2005eb,Afonso:2007zz,Dzhunushaliev:2009dt} and the Palatini $f(R)$ theory \cite{Bazeia:2014poa,Gu:2014ssa}. As  $f(T)$ gravity theory came up as an alternative to dark energy for the explanation of the acceleration of the universe \cite{Bengochea:2008gz}, it was then widely investigated \cite{Cai:2015emx,Ferraro:2006jd,Bamba:2012vg,Fiorini:2013hva,Geng:2014nfa}.   Braneworld models in $f(T)$ theories were studied in Refs.~\cite{Yang:2012hu,Menezes:2014bta,Guo:2015qbt}. In the previous works \cite{Yang:2012hu,Menezes:2014bta}, the solutions of braneworld scenarios in $f(T)$ gravity with the form of $f(T)=T+\alpha T^n $ were investigated by the first-order formalism, i.e., the superpotential way. Besides, the split of the brane in $f(T)$ gravity was given in Ref. \cite{Yang:2012hu}. Furthermore, the tensor perturbation of the braneworld was also studied in Ref. \cite{Guo:2015qbt} and it was shown that the solutions to the $f(T)$ braneworld were stable. The localization of matter fields was also investigated in Ref. \cite{Yang:2012hu}.

On the other hand, there are also braneworld scenarios in which the kinetic terms of background scalar fields are of non-canonical form, i.e., the $K$-fields. The $K$-field theory was first proposed as a new mechanism of inflation in cosmology \cite{Garriga:1999vw,ArmendarizPicon:1999rj,ArmendarizPicon:2000ah}. The lagrangian of a $K$-field can be written as $\mathcal{L}=\mathcal{F}(X, \phi)-V(\phi)$, where $X=-\frac{1}{2}g^{MN}\partial_M\phi\partial_N\phi$ is the kinetic term of the scalar field and $\mathcal{F}(X,\phi)$ is an arbitrary function of $X$ and $\phi$. The $K$-field theory was then applied to the thick branworld models \cite{Bazeia:2003cv,Koley:2005nv,Adam:2008ck,Bazeia:2008zx,Liu:2009ega,Castro:2010uj}. The analytical solutions of the thick $K$-brane models with two specific cases $\mathcal{L}=X-\alpha X^2-V(\phi)$  and $\mathcal{L}=-X^2/2-V(\phi)$ were given by the first-order formalism in Ref.~\cite{Bazeia:2008zx} with the perturbative procedure and in Ref.~\cite{Zhong:2013xga} with the non-perturbative procedure. Furthermore, the stability of thick $K$-brane was  discussed in Ref. \cite{Zhong:2012nt} and the trapping of bulk  fermions was also discussed in Ref. \cite{Castro:2010uj}.

Besides a mechanism of
inflation, the K-fields also can be a dynamical dark energy model \cite{Akhoury:2008dv,Armendariz-Picon2005,Bertacca:2007fc}.
 As a scalar field pervading the universe, from the effective field theory view
point, the K-fields must interact with the matter that is present\cite{Gauthier:2010zza}. One of the interesting cases is the interaction with the neutrino, which undergoes flavor oscillations when traveling through the K-field background. By introducing some certain types of CPT violating terms in the neutrino
action, it is possible to provide a unified explanation for all the existing neutrino data\cite{Katori:2006mz,Athanassopoulos:1997pv,Gauthier:2010zza}. One of the ways that may give a CPT violating term is considering the effect of torsion\cite{Gauthier:2010zza}. Additionally, there is also work about torsion induced neutrino oscillations\cite{DeSabbata:1981ek}. So torsion in the K-fields is one interesting possibility to the interaction between dark matter and neutrino. We are especially interested in the higher dimensional aspects of this model and corresponding braneworld scenario.

In this paper, we will consider the braneworld model of $f(T)$ gravity with $K$-fields. The forms of $f(T)$ will be more general. Instead of adopting the usual superpotential method, we develop a new way to give the solutions and we get some specific solutions for different
forms of $f(T)$ and $K$-fields. Based on these solutions, we can then study the energy density of the background fields and analyse the structure of the branes. Moreover, we find that by adding
terms in the polynomial form of $f(T)$ the number of split
sub-branes will increase and this gives us a mechanism to
generate sub-branes.

In the next section, we will give a review of the braneworld scenario in $f(T)$ gravity briefly. And we will introduce the new method to construct the braneworld with $K$-fields in $f(T)$ gravity. In  Sec. \ref{secIII}, we get the solutions for different forms of $f(T)$ and $K$-field. Specially, the form of $f(T)$ is taken as a polynomial of $T$ with arbitrary terms and the Lagrangian of background fields also contains arbitrary terms of polynomials of the canonical kinetic term. Based on these solutions, in Sec. \ref{secIV}, we study the energy density of matter fields along the extra dimension, which gives us the split of the brane. In Sec. \ref{secV}, we study the tensor perturbation of our model and recover the four-dimensional effective gravity. Finally, a brief summary is given in Sec. \ref{secVI}.

\section{Background and $f(T)$ Braneworld}
In  teleparallel
gravity, the dynamical fields are vielbein fields $e_A(x^M)$ associated with each point in the manifold with spacetime coordinates $x^M$. In the case of the braneworld scenario considered here, the spacetime is a five-dimensional manifold. We use the indices $A, B, C,\cdots$ and $M, N, P,\cdots$ which run from 0 to 4 to label the coordinates of tangent space and spacetime, respectively. Furthermore, the relation of the metric and vielbein is given by
\begin{equation}
g_{MN}=e^A_M e^B_N \eta_{AB},
\end{equation}
where $\eta_{AB}=\text{diag}(-1,1,1,1,1)$. And the Weitzenb\"{o}ck connection $\Gamma^P_{\ MN}$ is
\begin{equation}
\Gamma^P_{\ MN} \equiv e^P_B \partial_N e^B_M=-e^B_M \partial_N e^P_B.
\end{equation}
The torsion tensor is defined as
\begin{equation}
T^P_{\ MN} \equiv \Gamma^P_{\ MN}-\Gamma^P_{\ NM}=e^P_B(\partial_N e^B_M-\partial_M e^B_N).
\end{equation}
The difference between the Weitzenb\"{o}ck connection and the Levi-Civita connection $\tilde{\Gamma}^P_{\ MN}$ widely used in general relativity is defined as the contorsion tensor
\begin{eqnarray}
K^P_{\ MN} &\equiv& \Gamma^P_{\ MN}-\tilde{\Gamma}^P_{\ MN} \nonumber\\
  &=& \frac{1}{2}(T_{M \ N}^{\ \ P}+T_{N \ \ M}^{\ \ P}-T^P_{\ MN}).
\end{eqnarray}
From the torsion and contorsion tensors, the tensor $S^{PMN}$ can be defined as
\begin{equation}
S^{PMN} \equiv K^{MNP}-g^{PN}T^{QM}_{\ \ \ \ Q}+g^{PM}T^{QN}_{\ \ \ \ Q},
\end{equation}
and then the torsion scalar $T$ is given by
\begin{equation}
T \equiv \frac{1}{2}S^{PMN}T_{PMN}.
\end{equation}
Thus, one can write the Lagrangian of the teleparallel gravity as
\begin{equation}
L_T=-\frac{c^4 e}{16 \pi G}T,
\end{equation}
where $e$ is the determinant of the vielbein $e_A(x^M)$. As to $f(T)$ gravity in the braneworld scenario, the action is given by replacing $T$ with $f(T)$, which is an arbitrary function of $T$. Considering the matter field, the total action is
\begin{equation}\label{action}
S=-\frac{1}{4}\int d^5 x ~e f(T) +\int d^5 x \mathcal{L_M},
\end{equation}
where we have taken $\frac{c^4}{4\pi G}=1$ for convenience, and $\mathcal{L_M}$ is the Lagrangian of background matter fields. Instead of adopting a canonical form of matter fields, in our work, the kinetic term of a scalar field is replaced with an arbitrary function $P(X)$ of the canonical kinetic term of the scalar field $X$, i.e., the Lagrangian of the background matter is of the form
\begin{equation}
\mathcal{L_M}= e (P(X)-V(\phi)),
\end{equation}
where the canonical kinetic term is of the form  $X=-\frac{1}{2}g^{MN}\partial_M \phi\partial_N \phi$.

Varying the action (\ref{action})  with respect to the vielbein, we get the field equation
\begin{equation}\label{EOMF}
\begin{split}
&\frac{1}{4}g_{MN}f(T)+\frac{1}{2}f_T \Big[ g_{R M} S_P^{\ R Q} e_N^A\partial_Q e^P_A\\
&+e^{-1}g_{R M} \partial_Q(S_N^{\ R Q}e)- g_{R M} T^P_{\ QN} S^{\ QR}_P \Big]\\
&+\frac{1}{2}g_{R M} f_{TT} S_N^{\ R Q}\partial_Q T=\mathcal{T}_{MN},
\end{split}
\end{equation}
where $f_T$ and $f_{TT}$ represent respectively the derivative and the second-order derivative of $f(T)$ with respect to $T$: $f_T \equiv df/dT$ and $f_{TT} \equiv d^2f/dT^2$, and $\mathcal{T}_{MN}$ is the energy-momentum tensor defined as
\begin{equation}\label{energymomentumtensor}
\begin{split}
\mathcal{T}_{MN}&=e^{-1 }  g_{R M} e^A_N\frac{\delta \mathcal{S_M}}{\delta e^A_R}\\
&=g_{MN}[P(X)-V(\phi)]+P_X\partial_M \phi \partial_N \phi, \\
\end{split}
\end{equation}
where $\mathcal{S_M}$ is the action corresponding to the Lagrangian $\mathcal{L_M}$, and $P_X$ is the derivative of $P(X)$ with respect to $X$.

Next we consider the static flat braneworld scenario, for which the metric is
\begin{equation}
ds^2=e^{2A(y)}\eta_{\mu\nu}dx^\mu dx^\nu+dy^2,
\end{equation}
where the Greek indices run from 0 to 3 and $\eta_{\mu\nu}=\text{diag}(-1,1,1,1)$ is the four-dimensional Minkowski metric. Besides, $e^{2A(y)}$ is the warped factor and $y$ is the coordinate of the extra dimension. We can choose the vielbein in the form $e^A_M=\text{diag}(e^A,e^A,e^A,e^A,1)$, since it imposes no constraints on the function $f(T)$ and the torsion scalar $T$ \cite{Tamanini:2012hg,Ferraro:2011us}. Then the torsion scalar can be calculated as $T=-12 A'^2$. The prime denotes the derivative with respect to $y$ through the whole paper. Besides, we also require the scalar field  $\phi$ to depend on $y$ only. Then the equation of motion of the background field can be given by varying with respect to $\phi$:
\begin{equation}\label{scalarEOM}
\partial_y (e P_X \phi') = e { V}_{\phi}.
\end{equation}
Using the metric assumption above, it is easy to write the equations of motion as follows:
\begin{eqnarray}
\frac{1}{4}f+\frac{1}{2}f_T \big(3 A''&+&12A'^2\big)-36f_{TT}A'^2 A''\nonumber\\
&=&P -V,\label{EOM1}\\
\frac{1}{4}f+6 f_T A'^2&=& P-V+P_X\phi'^2,\label{EOM2}\\
4A' \phi' P_X+\partial_y(P_X \phi')&=& V_{\phi},\label{EOM3}
\end{eqnarray}
where Eqs. (\ref{EOM1}) and (\ref{EOM2}) are derived from the $\mu\nu$ components and $44$ component of Eq. (\ref{EOMF}) respectively, and Eq. (\ref{EOM3}) could be given by simplifying Eq. (\ref{scalarEOM}).


These are second-order differential equations and it is not easy to get analytic solutions in general case. The most popular method to give braneworld solutions is the superpotential way. And the superpotential way was used in braneworld scenario of $f(T)$ gravity with canonical kinetic term in Ref.~\cite{Yang:2012hu}. Besides, for the specific form of the non-canonical kinetic $P(X)=X+\alpha[(1+b X)^n-1]$, the solution was also given by the superpotential way \cite{Menezes:2014bta}. In both cases, only the form of $f(T)=T+\alpha T^n$ was considered. So it is natural to consider more general forms of $f(T)$ and $P(X)$.
To this end, we need to consider other method. Instead of using a superpotential approach to solve these equations as usual, in this paper, we develop a new approach to solve the equations by which we could get some new and interesting solutions easily.

Considering that only two of the three equations are independent, we will focus on Eqs. (\ref{EOM1}) and (\ref{EOM2}). Subtracting Eq. (\ref{EOM2}) from Eq. (\ref{EOM1}), we get
\begin{equation}
\frac{3}{2}f_T A''-36 f_{TT}(A')^2 A''=-P_X({ \phi'})^2,
\end{equation}
which can also be written as
\begin{equation}\label{algeeq}
\frac{3}{2}\partial_y\left(f_T A'\right)=-P_X({\phi'})^2.
\end{equation}
This equation gives a prescription to solve the system: noting that $A(y)$ and $T$ are functions of $y$ only, we can see that the left hand side of Eq. (\ref{algeeq}) only depends on $y$. As for the right hand side of the equation, it is a function of $\phi'$ only, because $P(X)$ is an arbitrary function of $X=-\frac{1}{2}g^{MN}\partial_M \phi\partial_N \phi = -\frac{1}{2} \phi'^2$. Regarding $\phi'$ as a new variable, we can get a polynomial equation or transcendental equation of $\phi'$ and $y$, which is easier to solve by some skills than a differential equation of $\phi$. After solving this polynomial equation or transcendental
equation, we can get the expression of $\phi'$. Then, by integrating $\phi'$ we at last get the solution of $\phi$. In the next section, we will give some specific solutions of the system with different functions $f(T)$ and $P(X)$.

\section{some specific solutions}\label{secIII}

In our work, we only consider the following warp factor solution
\begin{equation}
A(y)=-m \ln \left(\cosh (k y)\right)\label{warpfactora},
\end{equation}
where $k$ is an arbitrary constant with mass dimension one. It should be noted that the bulk spacetime is asymptotically $AdS_5$ at $|y|\rightarrow \infty$ and hence the braneworld considered in this paper is embedded in an $AdS_5$ spacetime. For the requirement that the warp factor is not divergent at the boundary of the extra dimension, we set $m>0$.

In the following part, we will consider two specific forms of $f(T)$:
\begin{equation}
  f(T)=T_0\left(e^{\frac{T}{T_0}}-1\right)
\end{equation}
and
\begin{equation}
 f(T)=\sum_{n=0}^{N}\frac{\alpha_{n}}{n+1} T^{n+1} +C,
\end{equation}
for which we have
\begin{equation}
  f_T(T)= e^{\frac{T}{T_0}}
\end{equation}
and
\begin{equation}
 f_T(T)=\sum_{n=0}^N \alpha_n T^n,
\end{equation}
respectively. Note that in the second case, $\alpha_n$'s are arbitrary constants with mass dimension $-2n$ to ensure that $f_T$ is dimensionless and $N$ could be any positive integer. So the second case can represent a wide class $f_T$. Next, we will give the solutions for both cases with different forms of  $P(X)$.

\subsection{$f_T=\exp({T}/{T_0})$}

For simplicity, we set $T_0=24 k^2 m^2$ throughout the paper. The left hand side of Eq. (\ref{algeeq}) gives
\begin{equation}\label{lhsof218}
\frac{3}{2}\partial_y\left(f_T A'\right)=-\frac{3}{2} e^{-\frac{1}{2} \tanh^2(k y)} k^2 m \cosh^{-4}(k y).
\end{equation}
This expression can be regarded as the product of an exponential function of $\tanh^2(k y)$ and a polynomial of $\tanh^2(k y)$ because $\cosh^{-2}(k y)=1-\tanh^2(k y)$. Firstly, if the right hand side of Eq. (\ref{algeeq}) can be written as some power functions of $X$, for instance, $X^n$, we can get $X$ as the $n$-th root of the right hand side of Eq. (\ref{lhsof218}). And if the form of $P(X)$ we adopt is pretty good, it could be easy to get the analytical solution of $\phi(y)$ by integrating $\phi'$. In the following, we'll give solutions in the case of $P(X)=X$ and $P(X)=C_1\sqrt{-X}$, where $C_1$ is a constant.

For $P(X)=X$, we can easily get the solution
\begin{eqnarray}
\phi(y) &=&   v_1 \,\text{erf}\Big(\frac{1}{2} \tanh(ky)\Big),\\
V(\phi)  &=&  \frac{3}{4} k^2 m\,  e^{-2\mathcal{F}^2(\phi) }
    \Big[ \left(1 - 4 \mathcal{F}^2(\phi) \right)^2 \nonumber \\
  &+&   8 m \left( e^{2 \mathcal{F}^2(\phi)}
                   - 4 \mathcal{F}^2(\phi) - 1
                 \right)
    \Big],
\end{eqnarray}
where $v_1=\sqrt{\frac{3 \pi m}{2}}$, $\mathcal{F}(\phi) = \text{erf}^{-1}({\phi}/{v})$ with $\text{erf}^{-1}(x)$ the inverse of the error function $\text{erf}(x)$.

For $P(X)=C_1\sqrt{-X}$, we have
\begin{eqnarray}
\phi(y)&=&-v_2\,  e^{-\frac{1}{2} \tanh ^2(k y)} \tanh (k y),\\
V(\phi)&=&6k^2 m^2\, e^{\frac{1}{2} \text{ProductLog}(-{\phi^2}/{v_2^2})}  \Big[ e^{-\frac{1}{2} \text{ProductLog}(-{\phi^2}/{v_2^2})} \nonumber\\ &&+\text{ProductLog}(-{\phi^2}/{v_2^2}) -1\Big],
\end{eqnarray}
where $v_2=\frac{3 k m }{\sqrt{2} C_1}$ and $\text{ProductLog}(x)$ gives the principal solution for $w$ in $x=w e^w$.

More complicatedly, if the right hand side of Eq. (\ref{algeeq}) could be written as the same form of the exponential function of $X$ times a polynomial of $X$, we are able to read $X$ off. Once again, if the forms of $P(X)$ are pretty good, we could get the analytical solutions. So next, we will give some examples of $P(X)$ as we considered above:
\begin{eqnarray}
 &&P_X=-{X_0 }{X}^{-1} e^{c\sqrt{-X}},\\
 &&P=\frac{m k^2 }{\phi_0^2} (X + \phi_0^2)\, e^{-\frac{1}{2} - \frac{X}{\phi_0^2}}  ,\\
 &&P=-3 m k^2\left( \sqrt{\frac{-2X}{\phi_0^2}}-2\right) e^{\sqrt{\frac{-X}{2\phi_0^2}}-\frac{1}{2}},
\end{eqnarray}
where $X_0$, $\phi_0$ and $c$ are constants. The corresponding solutions are given as follows.

For $P_X=-(X_0/X) \exp(c\sqrt{-X})$, the solution reads
\begin{eqnarray}
 \phi(y)=&&\frac{1}{c k}\sqrt{2} \ln \left(\frac{3 k^2 m}{2X_0}\right) \bigg[\texttt{Li}_2(-e^{-2 k y})\nonumber\\
 &&-k y (-k y-2+2 \ln 2)\nonumber\\
 &&+ 2 \tanh (k y) \bigg(\ln \big(\cosh^{-1}(k y)\big)-1 \bigg)\bigg],\\
V(y)=&&-6 m^2 k^2 \left( e^{-\frac{1}{2} \tanh^2(k y)} -1 \right)\nonumber\\
&& - 2 X_0 \texttt{Ei}\bigg(
\bigg|\bigg[\ln\frac{3 k^2 m}{2 X_0}+ 4 \ln \big (\cosh^{-1}(k y) \big)\nonumber\\
&& - \frac{1}{2} \tanh^2(k y)\bigg]^2/
c^2\bigg|\bigg)\nonumber\\
&&+ \frac{3}{2} m  k^2\, e^{-\frac{1}{2} \tanh^2(k y)} \cosh^{-4}(k y)\nonumber\\
&&-
 6 m^2 k^2 \, e^{-\frac{1}{2} \tanh^2(k y)} \tanh^2(k y),
\end{eqnarray}
where $\texttt{Li}_{2}(x)$ is the polylogarithm function of $x$ and $\texttt{Ei}(x)$ is the exponential integral function of $x$.

For $P(X)=(m k^2 /\phi_0^2) (X + \phi_0^2)\exp(-1/2 - X/\phi_0^2)$, we get the following $\phi(y)$ and $V(y)$:
\begin{eqnarray}
\phi(y)&=&\frac{2 \phi_0 }{k} \tan ^{-1}\left(\tanh \left(\frac{k y}{2}\right)\right),\\
V(\phi)&=&\frac{1}{2} m k^2 \, e^{-\frac{1}{2} \sin^2(\frac{k\phi}{\phi_0})}  \bigg((1 - 12 m) (1+\sin^2(\frac{k\phi}{\phi_0}))\nonumber\\
   && +
   12 m \, e^{\frac{1}{2} \sin^2(\frac{k\phi}{\phi_0})}  + 3 \cos^4(\frac{k\phi}{\phi_0}) \bigg).
\end{eqnarray}

For $P(X)=-3 m k^2\left( \sqrt{-\frac{2X}{\phi_0^2}}-2\right) \exp\left(\sqrt{-\frac{X}{2\phi_0^2}}-\frac{1}{2}\right)$, we get
\begin{eqnarray}
 \phi(y)&=&\frac{\phi_0}{k}\tanh (k y),\\
V(\phi)&=&\frac{3}{2} mk^2 \, e^{-\frac{1}{2}(\frac{k\phi}{\phi_0})^2}
   \bigg[4 +
   4 m\left( e^{\frac{1}{2} (\frac{k\phi}{\phi_0})^2}-2 \right) \nonumber \\
   &&+ ( 4 m-2) \left(1-(\frac{k\phi}{\phi_0})^2\right)\nonumber\\ &&+
   \left(1-(\frac{k\phi}{\phi_0})^2\right)^2\bigg].
\end{eqnarray}

\subsection{$f_T=\sum_{n=0}^N \alpha_n T^n$}

 For $f_T=\sum_{n=0}^N \alpha_n T^n$, the left hand side of Eq. (\ref{algeeq}) is given by
\begin{equation}
\begin{split}
\partial_y(\frac{3}{2}f_T A')=&\sum_{n=0}^N -\frac{\alpha_n}{2} 3^{1+n} 4^n m(1+2n)\bigg(-k^2 m^2 \big(1-\\
&\cosh^{-2}(ky)\big)\bigg)^n\cosh^{-2}(ky).
\end{split}
\end{equation}
Following the sprit in the last subsection, we consider the right hand side of this equation as a polynomial of $\cosh^{-2}(k y)$ again. Then comparing with the right hand side of Eq. (\ref{algeeq}), which is a polynomial of $X$, we can set $X \propto\cosh^{-2}(k y)$, i.e., $\phi' = \phi_0 \cosh^{-1}(k y)$, where $\phi_0$ is an arbitrary constant, and read off the corresponding $P(X)$. The solution is given by:
\begin{eqnarray}
\phi &=&\frac{2 \phi_0}{k} \tan ^{-1}\left(\tanh \left(\frac{k y}{2}\right)\right),\\
P&=&\sum _{n=0}^N -\alpha _n\beta^{(n)} \left(-\frac{k^2 m^2 }{\phi_0^2} \left(2 X+\phi_0^2\right) \right)^{n+1},\\
V&=&\sum_{n=0}^N \bigg[ -\alpha_n \beta^{(n)}
  \left(-k^2 m^2 \sin^2(\frac{k\phi}{\phi_0})\right)^{1+n}\nonumber\\
&&+ \frac{3}{2} \alpha_n m k^2 (2n+1){\bigg(-12k^2 m^2\sin^2(\frac{k\phi}{\phi_0})\bigg)^n\cos^{2}(\frac{k\phi}{\phi_0})}\nonumber\\
&&-6 12^n \alpha_n \left(-k^2 m^2 \sin^2(\frac{k\phi}{\phi_0})\right)^{n+1}\bigg]\nonumber\\
&&-\frac{1}{4}\sum_{n=1}^N \frac{\alpha_n}{n}\Big(-k^2 m^2 \sin^2(\frac{k\phi}{\phi_0})\Big)^n,
\end{eqnarray}
where $\beta^{(n)}=\frac{3^{n+1} 4^{n-1} (2 n+1)}{m(n+1)}$.

We should note that the solutions have a large dependence on the forms of potentials. In other words, our method works only for  some particular potential. This could be seen from our procedure. Our method is mainly based on solving the polynomial equation or transcendental equation gotten from Eq. (\ref{algeeq}). If we fix the forms of $f(T)$ and $P(X)$, the polynomial equation or transcendental equation gives us some special forms of $\phi$ with corresponding potential $V$ as the solution. Thus only the system with these potential could be solved in our way. As for other forms of potential, whether they could be solved needs further discussion.

The solutions and potentials are obtained in classical circumstances. Then another question arises naturally: if we consider quantum effects, is there any symmetry protecting these potentials from quantum fluctuations? The potential is usually assumed to emerge from the microscopic physics which leads to the braneworld model in a low energy limit \cite{DeWolfe:1999cp}. For the usual super potential way, the potential is the one in five-dimensional gauged supergravity with a $USp(8)$ invariance \cite{Freedman:1999gp,DeWolfe:1999cp}. But it's an interesting and open question if one can construct a supergravity theory that the supersymmetry conditions could give any potential desired \cite{DeWolfe:1999cp}. We are not able to answer this question and we focus on the classical aspects of the braneworld in this paper. It is still unknown if all our potentials possess some symmetries that could protect them all from quantum fluctuation.

\section{The split of the brane}\label{secIV}

In this section, we investigate the split of the brane. This phenomenon can be generated by a real scalar field \cite{Bazeia:2003qt}, two real scalar fields \cite{Bazeia:2003aw,deSouzaDutra:2008gm,Liu:2011ysa,Fu:2011pu} or a complex scalar field \cite{Campos:2001pr}. In Ref. \cite{Yang:2012hu}, the authors pointed out that the effect of torsion will influence the distribution of the energy density. As a consequence, the brane will have an incomplete split. While the authors of Ref. \cite{Yang:2012hu} only considered $f(T)=T+\alpha T^n$, we can see more complicate phenomena caused by the arbitrary function of $T$ in our work. From Eq. (\ref{EOMF}), we get the energy density $\rho(y)$:
\begin{eqnarray}
\rho(y)&=&T_{MN} U^M U^N\nonumber\\
&=&[P-V]g_{MN} U^M U^N+\frac{\partial P}{\partial X}\partial_M \phi \partial_N \phi U^M U^N\nonumber\\
&=&-P(X)+V(\phi)\nonumber\\
&=&-\frac{1}{4}f(T)-6 f_T A'^2-\partial_y(\frac{3}{2}f_T A').
\end{eqnarray}
It is easy to see that $\rho$ only depends on $f(T)$ and $A(y)$.  Then we will consider two different forms of $f(T)$ from the previous sections and get the corresponding energy densities $\rho(y)$ to investigate the split of brane.

\begin{figure}[!htb]
  \begin{center}
  \includegraphics[height=4cm,width=6cm]{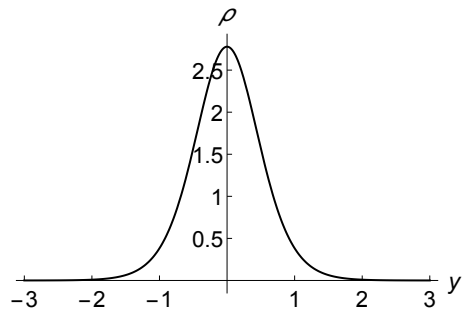}
  \caption{Plot of the energy density \eqref{EnergyDensity1} for $f_T=e^\frac{T}{T_0}$ with $k=m=1$.}\label{fig1}
  \end{center}
\end{figure}

For $f_T=e^\frac{T}{T_0}$, we have
\begin{equation}
\begin{split}
\rho&=\frac{3}{2} m k^2 \, e^{-\frac{1}{2} \tanh ^2(k y)} \\
&\times\bigg[4 m \left(e^{\frac{1}{2} \tanh ^2(k y)}-\tanh ^2(k y)-1\right)+\text{sech}^4(k y)\bigg].
 \label{EnergyDensity1}
\end{split}
\end{equation}
Considering that there could be an arbitrary constant, or so called cosmology constant in the potential $V$, we can separate it from $V$ by demanding $V$ to be zero at $y\rightarrow \infty$. Then we get the shape of the energy density (see Fig. \ref{fig1}).

From Fig. \ref{fig1}, we can easily see that the brane is not split and instead, it is localized near the zero point of the extra dimension.

\begin{figure}[!htb]
  \begin{center}
  \includegraphics[height=4cm,width=6cm]{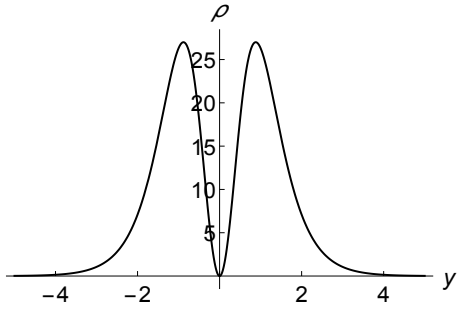}
  \end{center}
   \caption{Plot of the energy density \eqref{EnergyDensity2} for $f_T=\alpha_0+\alpha_1 T$ with $\alpha_0=-12$, $\alpha_1=-1$.}\label{fig2}
\end{figure}

Next, we will consider $f(T)$ satisfying the differential equation $f_T=\sum_{n=0}^N \alpha_n T^n$ which will result in a different energy density distribution along the extra dimension. We get the expression for $f(T)$ from the differential equation, i.e., $f(T)=\sum_{n=1}^{N+1}\frac{\alpha_{n-1}}{n} T^n +C$. Thus, the energy density is
\begin{equation}\label{rho}
  \begin{split}
   \rho=&\sum_{n=0}^N \alpha_n (-12)^n A'^{2n}(y)\\
   & \times \left[\left(\frac{3}{n+1}-6\right)A'^{2}(y)-\frac{3}{2} (2n+1)A''(y)\right].
   \end{split}
\end{equation}
To illustrate that the value of $N$ (or the number of  arbitrary constants $\alpha_n$'s in the expression of $f(T)$) could give us the split of brane, we now need to fix $N$ as some specific numbers and then give the calculations. Firstly, let us set $N=1$, i.e., $f_T=\alpha_0+\alpha_1 T$, then we get
\begin{eqnarray}
\rho
&=&54 \alpha_1 m^4 k^4 \tanh ^4(k y)  \nonumber \\
 &-&54 \alpha_1 m^3 k^4 \tanh ^2(k y) \text{sech}^2(k y) \nonumber \\
&-&3 \alpha_0 m^2 k^2 \tanh ^2(k y)+\frac{3}{2} \alpha_0 m k^2 \text{sech}^2(k y).~~~~\label{EnergyDensity2}
\end{eqnarray}
For simplicity, we will take $k=1$, $m=1$ in this section and the shape of the energy density for $N=1$ is shown in Fig. \ref{fig2}.
One can see that the brane is split into two sub-branes  which correspond to the two peaks in the energy density. Note that while $\alpha_0$, $\alpha_1$ are taken as some other combinations, there could be still one peak in the energy density which means that the brane is not split.  So in the following part, we will only display the graphs in which the brane have the maximum numbers of sub-branes.  Next, we'll show that the brane could split into more sub-branes with larger $N$.

\begin{figure}[!htb]
  \begin{center}
  \includegraphics[height=4cm,width=6cm]{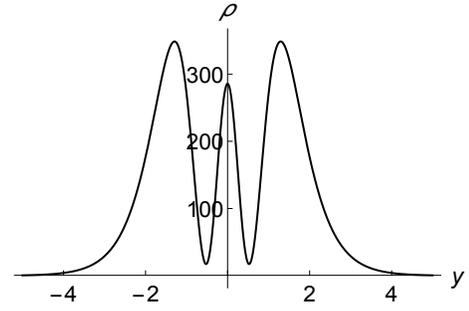}
  \end{center}
   \caption{Plot of the enrgy density \eqref{EnergyDensity3} for $f_T=\alpha_0+\alpha_1 T+\alpha_2 T^2$ with $\alpha_0=95.5$, $\alpha_1=40$, $\alpha_2=2.8$.}\label{fig3}
\end{figure}

Secondly, for $N=2$, i.e., $f_T=\alpha_0+\alpha_1 T+\alpha_2 T^2$, the energy density is
\begin{eqnarray}
\rho&=&-720 \alpha_2 m^6 k^6 \tanh ^6(k y) \nonumber \\
&+&1080 \alpha_2 m^5 k^6 \tanh ^4(k y) \text{sech}^2(k y) \nonumber\\
&+&54 \alpha_1 m^4 k^4 \tanh ^4(k y) \nonumber\\
&-&54 \alpha_1 m^3 k^4 \tanh ^2(k y) \text{sech}^2(k y) \nonumber\\
&-&3 \alpha_0 m^2 k^2 \tanh ^2(k y)+\frac{3}{2} \alpha_0 m k^2 \text{sech}^2(k y). \label{EnergyDensity3}
\end{eqnarray}
The shape is shown in Fig. \ref{fig3}. We can see that the brane is split into three sub-branes while $\alpha_0$, $\alpha_1$, $\alpha_2$ are taken as $95.5$, $40$, $2.8$, respectively.

\begin{figure}[!htb]
  \begin{center}
  \includegraphics[height=4cm,width=6cm]{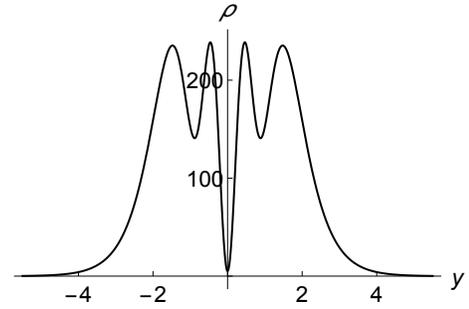}
  \end{center}
   \caption{Plot of the energy density \eqref{EnergyDensity4} for  $f_T=\alpha_0+\alpha_1 T+\alpha_2 T^2+\alpha_3 T^3$ with $\alpha_0= -116.4$, $\alpha_1= -47.2$, $\alpha_2=-7.48$, $\alpha_3= -0.371$.}\label{fig4}
\end{figure}

At last, we consider $N=3$, i.e., $f_T=\alpha_0+\alpha_1 T+\alpha_2 T^2+\alpha_3 T^3$, then
\begin{equation}
\begin{split}
\rho&=9072\alpha_3 k^8 m^8 \tanh^8(k y)\\
&-18144 \alpha_3 k^8 m^7 \tanh^6(k y)\text{sech}^2(k y)\\
&-720 \alpha_2 k^6 m^6 \tanh ^6(k y)\\
&+1080 \alpha_2 k^6 m^5 \tanh ^4(k y) \text{sech}^2(k y)\\
&+54 \alpha_1 k^4 m^4 \tanh ^4(k y)\\
&-54 \alpha_1 k^4 m^3 \tanh ^2(k y) \text{sech}^2(k y)\\
&-3 \alpha_0 k^2 m^2 \tanh ^2(k y)+\frac{3}{2} \alpha_0 k^2 m \text{sech}^2(k y).
\end{split}  \label{EnergyDensity4}
\end{equation}
It is shown in Fig. \ref{fig4}. As expected, there are four sub-brans.

For the general case of $f(T)=\sum_{n=1}^{N+1}\frac{\alpha_{n-1}}{n} T^n +C$, we could see that the maximum number of sub-branes increases with the number of terms in the polynomial expression of $f(T)$. This conclusion could be roughly seen from Eq. (\ref{rho}). Substituting Eq. \eqref{warpfactora} into Eq. \eqref{rho}, the right hand side of Eq. (\ref{rho}) becomes a polynomial of $\tanh^2(k y)$ whose highest order is $N+1$. So the distribution of energy density will be similar to a polynomial of $y^2$. Thus, we can get $N+1$ peaks by adjusting the values of $\alpha_n$'s. In other words, the more terms we adopt, the more sub-branes we could get.

\section{tensor perturbation and effective potential along the extra dimension}\label{secV}

In this section, we will investigate the linear tensor perturbation of the braneworld which satisfies the transverse-traceless condition. The perturbation of the vielbein fields is \cite{Guo:2015qbt}
\begin{equation}\label{vielbeinperturbation}
e^A_{\ M}=
\left[
\begin{matrix}
e^{A(y)}(\delta^a_{\ \mu}+h^a_{\ \mu})&0\\
0&1\\
\end{matrix}
\right],
\end{equation}
and the corresponding metric reads
\begin{equation}
g_{MN}=
\left[
\begin{matrix}
e^{2A(y)}(\eta_{\mu\nu}+\gamma_{\mu\nu})&0\\
0&1\\
\end{matrix}
\right],
\end{equation}
where
\begin{equation}
\gamma_{\mu\nu}=(\delta^a_{\ \mu}h^b_{ \ \nu}+\delta^b_{\ \nu}h^a_{\ \mu})\eta_{ab}.
\end{equation}
The transverse-traceless conditions are
\begin{equation}\label{TT}
\partial_\mu \gamma^{\mu\nu}=0=\eta^{\mu\nu}\gamma_{\mu\nu}.
\end{equation}
The perturbation of the torsion tensor, contortion tensor and e.t.c ($T^{\rho}_{\ \mu\nu}$, $K^{\rho}_{\ \mu\nu}$,$S^{\rho}_{\ \mu\nu}$ ) are given in Ref. \cite{Guo:2015qbt}.
The energy-momentum tensor is given as Eq. (\ref{energymomentumtensor}).
With the perturbation of the vielbein fields (\ref{vielbeinperturbation}) and the perturbation of the scalar field
\begin{equation}
\phi=\bar{\phi}+\tilde{\phi},
\end{equation}
where $\bar{\phi}$ is the background field and $\tilde{\phi}=\tilde{\phi}(x^\mu, y)$ is the perturbed field, we could get the perturbations of the $\mu\nu$ components of energy-momentum tensor
\begin{equation}
\delta T_{\mu\nu}=\Big(P_X(- \bar{\phi}'  \tilde{\phi}')-\frac{\partial V}{\partial \phi}\tilde{\phi}\Big) e^{2A} \eta_{\mu\nu}+(P-V)e^{2A}\gamma_{\mu\nu}.
\end{equation}
From Ref. \cite{Guo:2015qbt} the perturbation of the field equations (\ref{EOMF}) is given as
\begin{equation}
\begin{split}
\delta T_{MN}=&e^{-1}f_T\delta g_{NP}\partial_Q(eS_M^{\ PQ})+e^{-1}f_Tg_{NP}\partial_Q(eS_M^{\ PQ})\\
&+f_{TT}\delta S_{MN}^{\ \ \ \ Q}\partial_Q T-f_T\delta\tilde{\Gamma}^Q_{\ PM}S_{QN}^{\ \ \ P}\\
&-f_T\tilde{\Gamma}^Q_{\ PM}\delta S_{QN}^{\ \ \ P}+\frac{1}{4}\delta g_{MN}f(T).
\end{split}
\end{equation}
The $\mu\nu$ components give
\begin{equation}
\begin{split}
&\frac{1}{4}e^{2A}\gamma_{\mu\nu}+f_T e^{2A}\Big(6 A'^2\gamma_{\mu\nu}+\frac{3}{2}A''\gamma_{\mu\nu}\\
&-A'\gamma'_{\mu\nu}-\frac{1}{4}\gamma''_{\mu\nu}-\frac{1}{4}e^{-2A}\partial_\rho \partial^\rho \gamma_{\mu\nu}\Big)\\
& -f_{TT}e^{2A}\Big(36 A'^2 A''\gamma_{\mu\nu}-6 A' A'' \gamma'_{\mu\nu}\Big) \\
&=\Big(\frac{\partial P}{\partial X}(-\partial_y \bar{\phi} \partial_y \tilde{\phi})-\frac{\partial V}{\partial \phi}\tilde{\phi}\Big) e^{2A} \eta_{\mu\nu}\\
&+(P-V)e^{2A}\gamma_{\mu\nu}.\label{perturba}
\end{split}
\end{equation}
Subtracting the background equation (\ref{EOM1}) from Eq. \eqref{perturba}, we get
\begin{equation}
\begin{split}
&-f_T\Big(A'\gamma'_{\mu\nu}+\frac{1}{4}\gamma''_{\mu\nu}
          +\frac{1}{4}e^{-2A}\partial_\rho\partial^\rho\gamma_{\mu\nu}\Big)
+6f_{TT}A'A''\gamma'_{\mu\nu}\\
&=\Big(P_X(- \bar{\phi}' \tilde{\phi}')-\frac{\partial V}{\partial \phi}\tilde{\phi}\Big)\eta_{\mu\nu}.
\end{split}
\end{equation}
Contracting this equation with $\eta_{\mu\nu}$ and considering the transverse-traceless conditions  (\ref{TT}), we get
\begin{equation}
P_X(- \bar{\phi}' \tilde{\phi}')-\frac{\partial V}{\partial \phi}\tilde{\phi}=0,
\end{equation}
and
\begin{equation}
6f_{TT}A'A''\gamma'_{\mu\nu}
-f_T\Big(A'\gamma'_{\mu\nu}+\frac{1}{4}\gamma''_{\mu\nu}
         +\frac{1}{4}e^{-2A}\partial_\rho\partial^\rho\gamma_{\mu\nu}\Big)
=0,
\end{equation}
which is the same as Eq. (39) in Ref. \cite{Guo:2015qbt}.

As in Ref. \cite{Guo:2015qbt}, we could write the tensor perturbation equation as
\begin{equation}\label{tensorperturbation}
(\partial^2_z+2H\partial_z+\eta^{\mu\nu}\partial_\mu\partial_\nu)\gamma_{\mu\nu}=0,
\end{equation}
where
\begin{equation}
H=\frac{3}{2}\partial_zA+12e^{-2A}\Big((\partial_zA)^3-\partial^2_zA \partial_zA\Big)\frac{f_{TT}}{f_T},
\end{equation}
and $z$ is the conformal flat coordinate which is transformed from $y$ as
\begin{equation}
dz=e^{-A}dy.
\end{equation}
By introducing the Kaluza-Klein (KK) decomposition
\begin{equation}
\gamma_{\mu\nu}(x^\rho,z)=\epsilon_{\mu\nu}(x^\rho)F(z)\Psi(z),
\end{equation}
where $F(z)=e^{-\frac{3}{2}A(z)+\int K(z)dz}$ with
\begin{equation}\label{kz}
K(z)=12 e^{-2A}\Big(\partial^2_z A\partial_z A-(\partial_z A)^3\Big)\frac{f_{TT}}{f_T},
\end{equation}
we get two  equations from Eq. (\ref{tensorperturbation}). One is the equation for the four-dimensional KK gravitons $\epsilon_{\mu\nu}$:
\begin{equation}
(\eta^{\rho\sigma}\partial_\rho\partial_\sigma+\tilde m_n^2)\epsilon_{\mu\nu}(x^\rho)=0,
\end{equation}
and the other is the Schr\"{o}dinger-like equation for the extra-dimensional profile:
\begin{equation}\label{schrodingerlike}
\left(-\partial^2_z+U(z)\right)\Psi=\tilde m_n^2\Psi,
\end{equation}
where $\tilde m_n$ is the mass of the KK graviton and  $U(z)$ is the effective potential
\begin{equation}
U(z)=\partial_z H+H^2.
\end{equation}
The schrodinger-like equation (\ref{schrodingerlike}) can be factorized as
\begin{equation}
(\partial_z+H)(-\partial_z+H)\Psi=\tilde m_n^2\Psi,
\end{equation}
which means that $\tilde m_n^2>0$, i.e., any brane solution of $f(T)$ gravity theory with noncanonical scalar fields of the form $P(X)$ is stable under the transverse-traceless tensor perturbation.

Then we will consider the localization of the zero mode of graviton. As in Ref. \cite{Guo:2015qbt}, the zero mode of graviton is
\begin{equation}
\Psi_0=N_0 e^{\frac{3}{2}A-\int K(z)dz},
\end{equation}
where $N_0$ is the normalization coefficient.
The localization of the zero mode requires that
\begin{equation}\label{int}
\int \Psi_0^2dz=\int N_0^2e^{3A}e^{-2\int Kdz}dz<\infty.
\end{equation}

Substituting the expression of $K(z)$ (\ref{kz}) into Eq. \eqref{int}, we get
\begin{equation}
\int \Psi_0^2 dz=\int N^2_0 \cosh^{-2m}(k y) |f_T| dy.
\end{equation}
In this paper, the forms of $f_T$ are considered to be a polynomial of $T$ or an exponential function of $T$, so the integrand is not divergent. Thus, to check whether the requirement (\ref{int}) is satisfied, we only need to consider the asymptotic behavior of the integral when $y$ approaches to infinity. Thus, we can replace $\cosh^{-2m}(k y)$ with $e^{-2m |k y|}$ and get the integral at infinity
\begin{equation}
\int_\infty \Psi_0^2 dz=\int_\infty N^2_0 e^{-2m |k y|} |f_T| dy.
\end{equation}
Note that, $T$ is of the form $-12k^2m^2\tanh^2(ky)$ and $f_T$ is a polynomial of $\tanh^2(ky)$ or an exponential function of $\tanh^2(ky)$ in this paper. For both cases, $|f_T|$ is finite and we can see that if and only if $m>0$ could the integrand be integrable at infinity, which is in coincidence with the requirement of an asymptotically $AdS_5$ spacetime. So we conclude that the zero mode of graviton could be localized.

Next, we will give the effective potential and zero mode of graviton for two specific cases $f_T=\alpha_0+\alpha_1 T^1+\alpha_2 T^2$ and $f_T=e^{T/T_0}$.

\begin{figure}[!htb]
  \begin{center}
  \subfigure[The effective potential]{\label{effectivepotential1}
  \includegraphics[height=2.6cm,width=4cm]{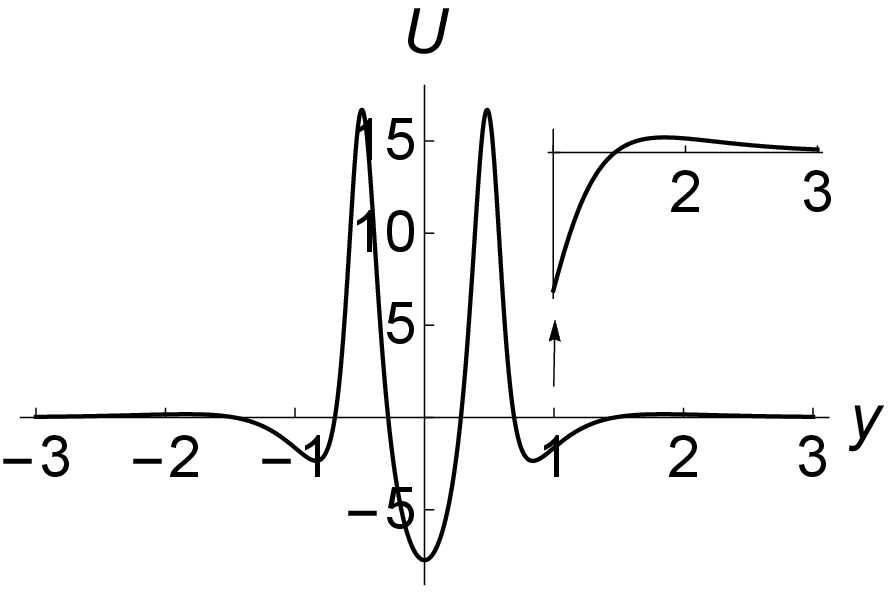}}
  \subfigure[The zero mode]{\label{zeromode1}
  \includegraphics[height=2.6cm,width=4cm]{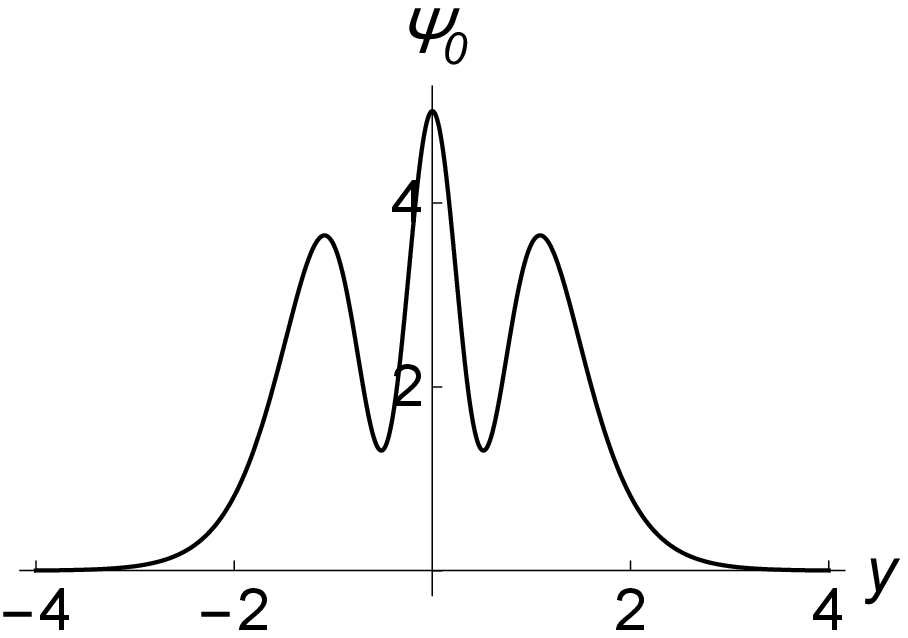}}
  \subfigure[The energy density]{\label{energydensity1}
  \includegraphics[height=2.6cm,width=4cm]{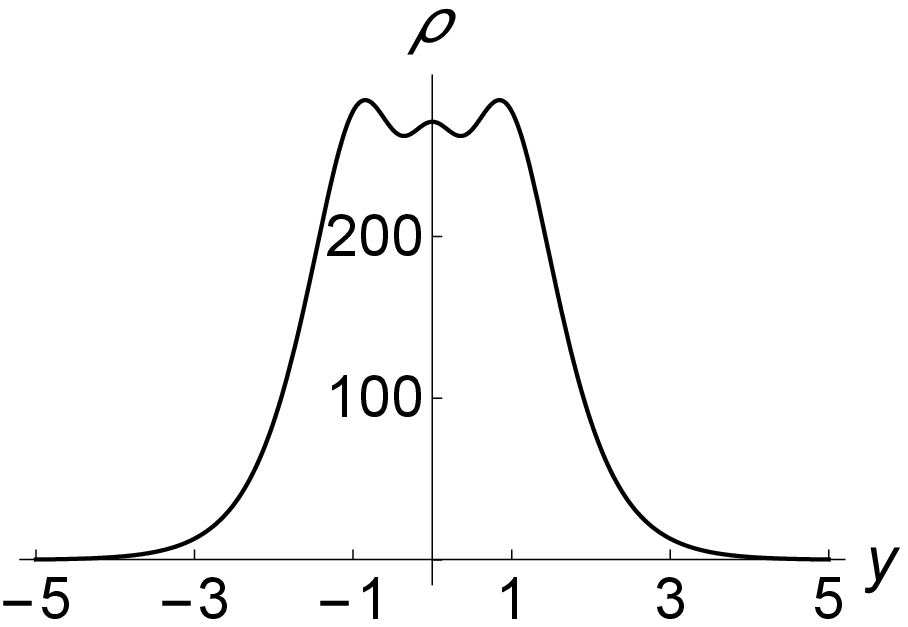}}
  \end{center}
   \caption{Plots of the effective potential, zero mode of graviton and energy density of the background field for $f_T=\alpha_0+\alpha_1 T^1+\alpha_2 T^2$ with $\alpha_0 = 5, \alpha_1 =2.6, \alpha_3 = 0.54$.}\label{fig6}
\end{figure}

For $f_T=\alpha_0+\alpha_1 T+\alpha_2 T^2$, the effective potential can be read as
\begin{widetext}
\begin{equation}
\begin{split}
U=&\cosh^{-2(2+m)}(k y)\bigg[-3k^2 m \alpha_0\bigg(8\alpha_0\cosh^2(k y)+32k^2m\alpha_1\big(2-2(2+7m)\sinh^2(k y)+15m^2\sinh^4(k y)\big)\\
&-5m \alpha_0\sinh^2(2k y)\bigg)+576k^6m^4\bigg(24\alpha_0\alpha_2+(\alpha_1^2+2\alpha_0\alpha_2)\sinh^2(k y)(-8-22m+15m^2\sinh^2(k y))\bigg)\tanh^2(k y)\\
&-41472k^8m^6\alpha_1\alpha_2\bigg(2-2(2+5m)\sinh^2(k y)+5m^2\sinh^4(k y)\bigg)\tanh^4(k y)\\
&+82944k^{10}m^8\alpha_2^2\bigg(8-2(8+19m)\sinh^2(k y)+15m^2\sinh^4(k y)\bigg)\tanh^6(k y)\bigg]\\
&\bigg/\bigg[16\bigg(\alpha_0-12k^2m^2\alpha_1\tanh^2(k y)+144k^4m^4\alpha_2\tanh^4(k y)\bigg)^2\bigg],
\end{split}
\end{equation}
\end{widetext}
and the zero mode of graviton is
\begin{equation}
\begin{split}
\Psi_0(y)=&N_0\cosh^{-\frac{3}{2}m}(k y) \Big|\alpha_0-12k^2m^2\alpha_1\tanh^2(k y)\\
&+144k^4m^4\alpha_2\tanh^4(k y)\Big|.
\end{split}
\end{equation}
The plots of the effective potential and zero mode are shown in Figs. \ref{effectivepotential1} and \ref{zeromode1}, where we have taken $k=m=1$.
These two figures show that the effective potential has three wells so the corresponding zero mode has three peaks. Compared with the energy density of the background field (see Fig. \ref{energydensity1}), we can see that the distribution of the zero mode of graviton is similar to that of the energy density, which implies that the zero mode could be localized near the brane. So we conclude that the split of the brane will cause the split of the zero mode of graviton.

\begin{figure}[!htb]
  \begin{center}
  \subfigure[The effective potential]{\includegraphics[height=2.6cm,width=4cm]{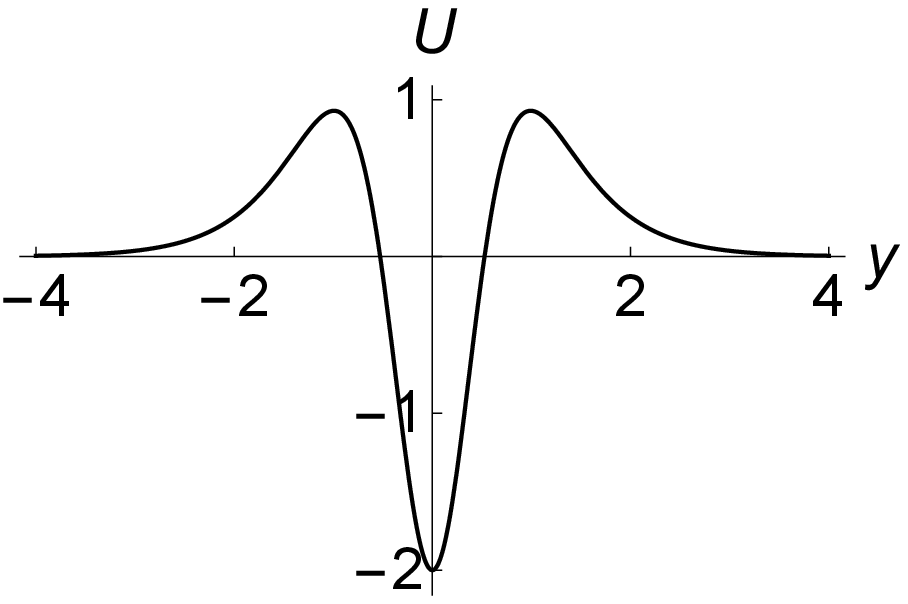}}
  \subfigure[The zero mode]{\includegraphics[height=2.6cm,width=4cm]{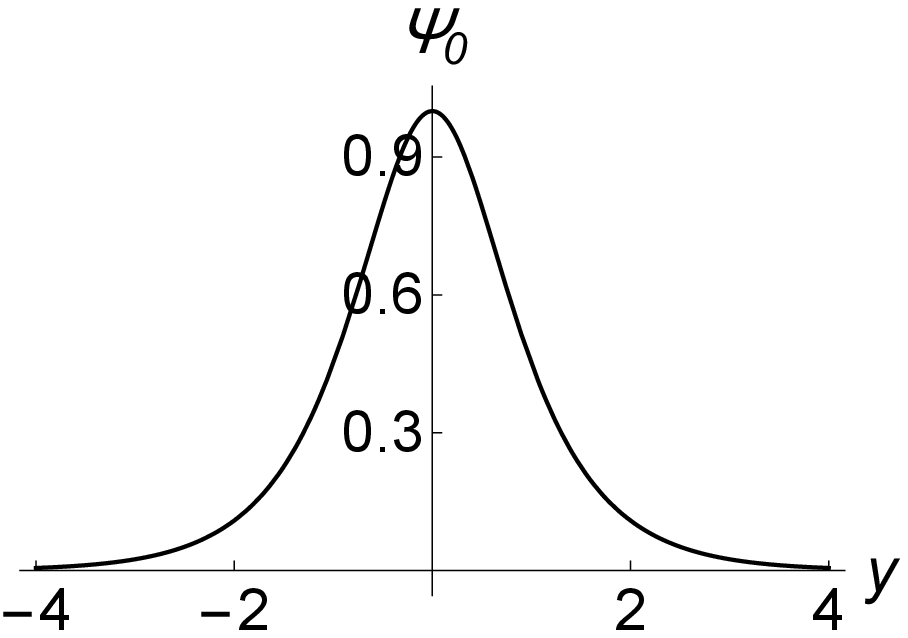}}
  \end{center}
   \caption{Plots of the effective potential and the zero mode of graviton for $f_T=e^{T/T_0}$.}\label{fig5}
\end{figure}

For $f_T=e^{T/T_0}$, we get
\begin{equation}
\begin{split}
U&= \frac{1}{4}k^2  \text{sech}^{2m+2}(k y)
   \Big( \text{sech}^2(k y)\tanh^2(k y)  \\
&+ (8m-6)\text{sech}^2(k y)
 + {15}m^2-6m+4 \Big),
\end{split}
\end{equation}
the corresponding zero mode of graviton is
\begin{equation}
\Psi_0(y)=N_0e^{-\frac{1}{4}\tanh^2(k y)}\cosh^{-3m/2}(k y).
\end{equation}
The corresponding figure for $k=m=1$ is shown in Fig. \ref{fig5}.
We can see that the effective potential is volcano-like and the zero mode is localized near the brane, this is similar to that of general relativity.

\section{Conclusion}\label{secVI}

In this paper, we developed a new method for finding solutions of the braneworld scenario in $f(T)$ gravity theory with K-fields. Following our method, we found several solutions to the cases that $f(T)$ takes the forms of $f(T)=T_0(e^{\frac{T}{T_0}}-1)$ and $f(T)=\sum_{n=1}^{N+1}\frac{\alpha_{n-1}}{n} T^n +C$. Then based on our solutions, we studied the distribution of the corresponding energy density along the extra dimension. The results shows that the polynomial form of the $f(T)$ will cause split of the brane. Next, we considered the stability of our solutions by investigating the linear tensor perturbation of the vielbein. And we concluded that our solutions are stable. Finally, we demonstrated that the zero mode of graviton could be localized for both forms of $f_T$. In addition, we calculated the effective potential of the KK modes of graviton along the extra dimension and gave the zero mode of graviton. ~~~~~

The brane we studied in this paper is flat, however, the cosmology constant on the brane can also be non-vanishing. Whether the method we developed in this paper is still valid for dS thick brane or AdS thick brane can be studied further. We will investigate this in the future.

\section{ACKNOWLEDGMENTS}

{{We thank the referee for his/her comments which improve this paper.}}
This work was supported by the National Natural Science Foundation of China (Grants Nos. 11875151, 11522541, and 11705070) and the Fundamental Research Funds for the Central Universities (Grants No. lzujbky-2018-k11).


\end{document}